# Multi-Scale Graph Theoretical Analysis of Resting-State fMRI for Classification of Alzheimer's Disease, Mild Cognitive Impairment, and Healthy Controls

Ali Khazaee[1], Abdolreza Mohammadi[1], Ruairi O'Reilly[2]

[1]Department of Electrical Engineering, University of Bojnord

[2]Department of Computer Science, Munster Technological University

## Abstract

Alzheimer's disease (AD) is a neurodegenerative disorder marked by memory loss and cognitive decline, making early detection vital for timely intervention. However, early diagnosis is challenging due to the heterogeneous presentation of symptoms. Resting-state functional magnetic resonance imaging (rs-fMRI) captures spontaneous brain activity and functional connectivity, which are known to be disrupted in AD and mild cognitive impairment (MCI). Traditional methods, such as Pearson's correlation, have been used to calculate association matrices, but these approaches often overlook the dynamic and non-stationary nature of brain activity. In this study, we introduce a novel method that integrates discrete wavelet transform (DWT) and graph theory to model the dynamic behavior of brain networks. Our approach captures the time-frequency representation of brain activity, allowing for a more nuanced analysis of the underlying network dynamics. Machine learning was employed to automate the discrimination of different stages of AD based on learned patterns from brain network at different frequency bands. We applied our method to a dataset of rs-fMRI images from the Alzheimer's Disease Neuroimaging Initiative (ADNI) database, demonstrating its potential as an early diagnostic tool for AD and for monitoring disease progression. Our statistical analysis identifies specific brain regions and connections that are affected in AD and MCI, at different frequency bands, offering deeper insights into the disease's impact on brain function.

# 1- Introduction

Alzheimer's disease (AD) is a progressive neurodegenerative disorder that is characterized by memory loss and cognitive decline. Mild cognitive impairment (MCI) is an intermediate stage before early AD and individuals with MCI are at a heightened risk of progressing to AD [1]. Currently, there is no known cure for AD, but various interventions, both pharmacological and nonpharmacological, have shown promise in slowing down its advancement, particularly when detected early [2]. Therefore, accurate prediction of AD progression and differentiation between its stages are crucial for timely medical intervention.

Neuroimaging has had significant implications for the field of neuroscience [3]. It has provided researchers with the ability to visualize brain activity, map neural networks, and investigate the underlying mechanisms of neurological disorders. Among neuroimaging techniques, Magnetic Resonance Imaging (MRI) and Functional Magnetic Resonance Imaging (fMRI) play pivotal roles in the automatic classification of AD. While MRI focuses on anatomical structure, revealing brain regions affected by cortical atrophy and amyloid plaques, fMRI measures changes in blood flow and oxygenation, providing insights into the brain's functional activity. Specifically, resting-state fMRI (rs-fMRI) captures spontaneous brain activity and functional connectivity, which are disrupted in AD and MCI patients [4, 5].

Most existing studies have used Pearson's correlation to calculate functional connectivity matrices, but this method fails to account for the dynamic nature of brain activity, resulting in a static representation that does not accurately reflect brain function. Some studies proposed sliding window methods to overcome this problem [6], but these approaches have limitations, such as the use of fixed window sizes and the introduction of redundant information, which can bias the analysis and increase computational complexity. These drawbacks highlight the need for more sophisticated methods, such as wavelet analysis, which offers superior time-frequency localization and noise reduction. The wavelet transform, a mathematical tool for decomposing signals into different frequency bands, has demonstrated its utility in various real-world applications. Despite the great potential of wavelet transform, it attracted little attention in neuroimaging. Recent studies have begun to explore its application, showing promising results in disorders like autism [7], depression [8], ADHD [9], and Parkinson [10]. AD and MCI also have been investigated using wavelet transform in a few studies [11-13]. These studies suggest that wavelet-based methods can effectively capture the complex dynamics of brain activity.

Our proposed method builds on this foundation by integrating discrete wavelet transform (DWT) with graph theory to model the dynamics of brain networks. Recent studies have explored the utility of graph-theoretical measures and machine learning in classifying Alzheimer's disease stages using resting-state fMRI [5, 14-17]. For example, Zamani et al. [15] employed an evolutionary optimization framework to select discriminative graph features for distinguishing early-MCI from healthy controls. Alorf et al. [16] applied deep learning to correlation-based connectivity matrices for multi-label classification of AD stages. Structural and functional network topologies were jointly analyzed by Lin et al. [17], revealing widespread alterations in AD. While these studies demonstrated the potential of graph features and deep learning, our approach extends this line of work by incorporating multi-scale signal decomposition via DWT to capture frequency-specific dynamics prior to graph construction. This added dimension enables a more detailed characterization of disease-related alterations across temporal scales and capturing the brain's dynamic behavior across different frequency bands.

The novelty of our method lies in its multi-resolution representation of the dynamic brain network, which enhances the ability to discriminate between different stages of AD. By applying DWT, we analyze brain signals at multiple frequency bands, offering a comprehensive view of functional connectivity across different temporal scales. This multi-resolution approach enhances the ability to detect subtle changes and patterns that may indicate different stages of AD. The integration of graph theory allows us to extract meaningful measures that reveal the underlying structure of the brain's functional connectivity at various frequencies. Additionally, machine learning algorithms are employed to automate the classification of subjects based on these graph measures, enabling accurate identification of AD and MCI stages.

## 2- Methods

### 2-1- Subjects, Data Acquisition and Preprocessing

This study will analyze rs-fMRI data of subjects from the Alzheimer's Disease Neuroimaging Initiative (ADNI) database (Table S1). The ADNI data is available to authorized investigators through the Image Data Archive (IDA). The Healthy control (HC), patients with AD, and patients with MCI will be selected based on the age, gender, and cognitive scores such as Mini-Mental State Examination (MMSE) and Clinical Dementia Rating (CDR) which indicate mild or moderate dementia. Functional and structural MRI data are collected according to the ADNI acquisition protocol using three tesla (3T) scanner [18]. The rs-fMRI data in each subject consisted of 140 functional volumes and acquired with following parameters: repetition time (TR) = 3000 ms; echo time (TE) = 30 ms; flip angle = 80º; slice thickness=3.313 mm; and 48 slices.

The standard preprocessing steps of rs-fMRI data will be performed. Briefly, the preprocessing steps are as follows: leaving the first few volumes (7 volumes) of the functional images for signal equilibrium and participant's adaptation to the circumstances; slice-timing correction to the last slice; realignment for head movement compensation by using a six-parameter rigid-body spatial transformation ; normalization to the Montreal Neurological Institute (MNI) space; resampling to 3-mm isotropic voxels; detrending; smoothing using a Gaussian filter with full width at half maximum (FWHM) of 4 mm; and band-pass filtering (0.01-0.08 Hz). During the realignment step, four subjects (two HC, one MCI, and one AD) who exhibited head motions more than 2.5 mm of displacement and 2.5 degrees of rotation in any direction, were excluded. The whole-brain signal is removed by a multiple linear regression analysis to reduce the effect of the physiological artifacts [19]. To reduce the effects of motion and non-neuronal blood oxygen level-dependent (BOLD) fluctuations, head motion as well as the cerebrospinal fluid (CSF) and white matter signals are removed as nuisance covariates [19, 20]. Preprocessing is carried out using the Data Processing Assistant for Resting-State fMRI (DPARSF) toolbox [21] and the SPM (Statistical Parametric Mapping) package (http://www.fil.ion.ucl.ac.uk/spm). The final dataset included 43 HC, 88 MCI, and 33 AD subjects.

**2-2- Brain network analysis**

**2-2-1- Time series extraction**

After preprocessing of the rs-fMRI images, the rs-fMRI signals of each voxel are extracted. The next step is to analyze these signals to identify patterns of functional connectivity between different brain regions. Brain parcellation is used for dividing the brain into distinct regions or parcels and use them as region of interests (ROIs). There are two main types of brain parcellation methods. Anatomical methods such as Automated Anatomical Labeling (AAL) and Talairach atlases are based on the physical structure of the brain. These methods use features such as cytoarchitecture, sulci, and gyri to define brain regions. Functional methods such as independent component analysis (ICA) are based on the functional activity of the brain. As we are studying brain function by rs-fMRI data, predefined functional brain atlases such as the 264 putative functional area atlas [22] are employed to parcellate the brain. The time series of voxels within each of ROI are averaged to generate a representative signal for each ROI.

### 2-2-2- Wavelet analysis

The DWT is a time-frequency analysis method that decomposes a signal into a series of wavelet coefficients. Wavelets are oscillatory functions that have a localized frequency content. This means that they can be used to represent signals that have a specific frequency content. Following standard preprocessing, including band-pass filtering between 0.01 and 0.08 Hz, DWT is applied to the rs-fMRI signals of each ROI to decompose them into frequency sub-bands across five levels. Given the sampling rate (TR = 3000 ms), the Nyquist frequency is approximately 0.167 Hz, and the 0.01–0.08 Hz band passed during preprocessing occupies the lower quarter of the usable frequency spectrum. Each DWT level corresponds to a dyadic sub-band (i.e., frequency halving per level), but due to the prior filtering, only DWT Levels 3 to 5 fall entirely or mostly within the preserved frequency range. Specifically, Level 5 primarily captures fluctuations around 0.01–0.02 Hz, which are associated with long-range, stable functional connectivity patterns. While higher-frequency DWT levels (Levels 1–2) may be affected by filtering and thus less reliable, they were included for completeness of the multi-scale analysis and to assess the potential discriminatory value of higher temporal dynamics.

Twenty different wavelet types spanning the Daubechies, Coiflet, Symlet, Biorthogonal, and Fejér-Korovkin families is tested. These wavelets were selected based on commonly cited criteria for biomedical signal processing, including compact support, orthogonality, smoothness, and symmetry, which help balance time-frequency localization with minimal phase distortion. This variety allowed us to explore how wavelet filter characteristics influence the decomposition of rs-fMRI signals and affect downstream network measures and classification performance. Association between the extracted rs-fMRI signals of ROIs is calculated using Pearson's correlation. This gives the original connectivity matrix. This procedure is repeated for decomposed rs-fMRI signals at different levels and result in frequency sub-band connectivity matrices.

**2-2-3- Thresholding**

As stated above, the original and frequency sub-band association matrices are constructed by calculation of Pearson's correlation of the corresponding signals of ROIs. The resulted association matrices are weighted dense connectivity matrices. For further analysis, we converted the weighted dense connectivity to sparse ones by thresholding them at an optimal threshold value. A detailed description of the method used to determine this threshold is provided in the Supplementary Materials (see S1). The binary undirected connectivity matrices for each subject are calculated by thresholding the connectivity matrices and setting their diagonal elements to zero.

**2-2-4- Computation of graph measures**

The connectivity matrices obtained for each subject were used to construct brain networks and calculate a comprehensive set of graph theory-based measures using the Brain Connectivity

Toolbox (BCT), a widely adopted software package for the analysis of structural and functional brain networks [23]. These measures provide insight into the organization, integration, and resilience of brain networks, particularly across different time-frequency scales.

The computed graph measures were categorized as follows:

- Local/Nodal Measures:
  These measures were computed for each of the 264 brain regions (nodes) and characterize properties at the node level. Node strength is the sum of weights of a node's connections. Local efficiency measures the fault tolerance of the system by evaluating communication efficiency in the node's immediate neighborhood. The ratio of local to global efficiency captures the balance between localized and distributed processing. Degree represents the number of direct connections a node has. Participation coefficient quantifies how evenly a node's connections are distributed across different network modules , while the diversity coefficient reflects the heterogeneity of a node's intermodular connectivity [23]. Betweenness centrality counts the number of shortest paths that pass through a node, indicating its role as a bridge. K-coreness centrality evaluates the maximal subset of nodes that are mutually interconnected. Subgraph centrality quantifies a node's involvement in all possible subgraphs. Eigenvector centrality indicates influence of a node based on the centrality of its neighbors, and PageRank centrality reflects the importance of a node considering both quantity and quality of its connections [5].

- Global Measures:
  These metrics reflect the overall structure and efficiency of the network. Characteristic path length represents the average shortest path between all node pairs. Global efficiency, the inverse of the average path length, indicates the network's capacity for parallel information transfer. Assortativity quantifies network resilience by measuring the tendency of nodes to connect with other nodes of similar degree. Average clustering coefficient reflects the overall network segregation and is computed by averaging the clustering coefficients of all individual nodes, where each node's clustering coefficient quantifies the likelihood that its neighbors are also interconnected. Small-Worldness is the ratio of clustering to path length, indicating the balance between local specialization and global integration in the network [5].

All graph measures, except for node strength, were computed using binary adjacency matrices derived by thresholding the connectivity matrices at the optimal proportional threshold. Node strength was calculated from the original weighted connectivity matrices, prior to thresholding. The final feature vector for each subject at each level of wavelet decomposition consisted of 2,909 features, including 2,904 local features (11 measures × 264 ROIs) and 5 global measures. These high-dimensional feature vectors were then used for classification of subjects into AD, MCI, and HC groups.

**2-3- Statistical Analysis Using NBS**

To determine the altered connectivity networks in AD and MCI patients, raw connectivity matrices were analyzed using the Network-Based Statistic (NBS) method [24]. The NBS was employed to identify statistically significant subnetworks of altered connectivity. This approach operates on a graph representation of the brain, where brain regions are nodes and their pairwise connections are links. Unlike conventional methods that test each connection independently, the NBS enhances statistical power by identifying connected components (subnetworks) of suprathreshold links. A detailed description of the NBS method is provided in the Supplementary Materials (S2).

Figure 1 illustrates the complete step-by-step workflow of this study, including data preprocessing, feature extraction, statistical analysis, and classification. After preprocessing of the rs-fMRI data, the 264 putative functional areas atlas was employed to parcellate the brain. Then, the signals of all voxels within each region were averaged to produce 264 signals for each subject. DWT was applied to each of the 264 extracted signals, and the approximate coefficients were computed for decomposition levels 1 through 5. This procedure was repeated using 20 different wavelet types (db1, db2, …). As a result of this part of our proposed method, for each ROI, 100 DWT-decomposed signals (corresponding to the 20 different wavelet types at five levels of decompositions) and one original signal were calculated. Pearson's correlation was applied to each group of extracted signals to construct 264-by-264 connectivity matrices of the brain network graph. For each subject, 101 connectivity matrices (one from correlating the original time series and 100 from correlating the DWT-decomposed time series of 20 different wavelet types at five levels of decomposition) were produced. The resulting connectivity matrices were used as input to the NBS for statistical analysis.

**2-4- Feature Selection and Classification**

An efficient feature selection algorithm is the essential part of a machine learning approach in case of high dimensional datasets, such as rs-fMRI. A combination of the Fisher score and forward sequential feature selection (FSFS) was employed in this study. We used the support vector machine (SVM) and naïve Bayes classifier as the supervised machine learning algorithms. The classification algorithms were implemented in MATLAB (The Math Works, Natwick, MA) using LIBSVM software package (www.csie.ntu.edu.tw/~cjlin/libsvm/). We used a holdout cross-validation approach to train and test the classifiers. For each experiment, K = 23 subjects per group (HC, MCI, AD) were randomly selected for training, and the remaining subjects were used for testing. This value was chosen to ensure balanced training sets while preserving enough data for evaluation, particularly considering the smallest group size (AD with 34 subjects). Given the computational complexity of the analysis—spanning 20 wavelet types, 5 decomposition levels, and forward sequential feature selection (FSFS) applied to high-dimensional feature spaces—we did not perform repeated cross-validation at this stage. Instead, we reported the performance from a single holdout split per configuration. While this limits robustness to sampling variability, it enabled systematic exploration across a broad design space.

# 3- Results

## 3-1- Statistical analysis on raw connectivity matrix

Using the raw connectivity matrices of subjects in AD, MCI, and HC groups, permutation-based group comparisons (10,000 permutations) were performed, and FDR correction was applied to the resulting p-values to control for multiple comparisons. The NBS analysis provided a sparse matrix of connections (edges) with significant differences between groups. Figure 2 shows the results for the original signals. Similar figures (and corresponding sparse matrices of significant edges) were obtained by applying NBS analysis to the 100 connectivity matrices of different wavelet types and different levels of decomposition. Finally, we obtained 101 sparse matrices of connections with significant differences among the three groups of HC, MCI, and AD. We aimed to find the most frequent significant edges (connections) at each level of decomposition. To do this, we accumulated the disrupted connectivity matrices at each level of decomposition across all 20 wavelet types. As a result, one matrix was generated for each level of decomposition. Figure 3 and Figures S1-S5 show these results on brain surfaces using the BrainNet Viewer software package. Table S2 shows the details of disrupted ROIs and connections (nodes and edges). The maximum weight of each connection was 20, corresponding to edges that were found significant across all NBS analyses at level i (i = 1, 2, 3, 4, and 5) using all wavelet types (db1, db2, …). Table S3 shows the details of significant nodes and connections when the original rs-fMRI signals were used to construct connectivity matrices (previously visualized in Figure 2). It is noteworthy that this process results in only one sparse matrix of significant connections.

**3-2- Application of machine learning approaches to graph measures**

After the statistical analysis, we investigated whether graph measures could differentiate patients with AD and MCI from healthy subjects. To this end, we used the proposed method in Section 2-2-3 and found the optimal value of $P = 0.22$ for converting the raw connectivity matrices into the sparse networks. Then the graph measures were calculated for sparse brain networks. The calculated measures were used as discriminative features to classify three classes: healthy control subjects, patients with MCI, and patients with AD. Feature selection was performed on the extracted features to select the most informative features. Filter feature selection algorithms were used as a pre-processing step since they are simple and fast. The filter feature selection algorithm (i.e. Fisher algorithm in this study) sorts the features based on their discrimination ability. The best feature is placed at the top of the list with maximum discrimination. Then, we added the FSFS wrapper stage after the filter stage. The FSFS was applied to the top 600 features selected from the filter stage to identify the final feature subset. Since the typical goal of classification is to maximize the accuracy, the FSFS feature selection procedure performs a sequential search using the accuracy of the learning algorithm on each candidate feature subset. The training set was used to select the features and to fit the model, and the test set was used to evaluate the performance of the final selected features. This procedure was repeated for features extracted using graph measures of connectivity matrices of different wavelet families at different levels of decomposition. Tables S4-S7 show the detailed classification results using wavelet families of the Coiflet, Daubechies, Fejér-Korovkin, and Symlet, respectively. Figures 4 and S6 summarize the obtained results.

## 4- Discussion

An altered network with 17 nodes and 16 edges was found using NBS analysis on raw connectivity matrices obtained from original rs-fMRI signals (Figure 2). About half of the disrupted connections involved regions in the Default Mode Network (DMN), with additional connections between DMN and both the visual and dorsal attention networks.

An investigation of the resting-state networks corresponding to the altered nodes across different levels of wavelet decomposition (as shown in Table S2 and Figure 3) revealed consistent patterns of disruption, primarily in the visual, DMN, uncertain, sensory/somatomotor hand, and dorsal attention networks (Table S3). A detailed breakdown of the specific networks affected at each of the five decomposition levels is provided in the Supplementary Materials (see S3). These findings align with previous studies reporting disrupted connectivity between the temporal, parietal, and occipital regions [25], as well as impairments in DMN, sensory/somatomotor, dorsal attention, and visual networks in AD and MCI [26-28].

Analysis of frequently disrupted regions consistently implicated key nodes in the DMN (e.g., Left Angular Gyrus), temporal, and occipital lobes. These disruptions are clinically relevant, as these areas are associated with memory, language, semantic processing, and visuospatial function, which are commonly impaired in AD [29-31]. A detailed list of the most frequently affected regions and their associated functions is available in the Supplementary Materials (see S4).

Our findings of disrupted connectivity in the Default Mode Network and visual networks are consistent with clinical manifestations of AD and MCI, where memory and visuospatial impairments are prominent [25, 32]. The most frequently disrupted connection (edge) was between the left angular gyrus and left middle occipital gyrus, linking DMN and visual processing pathways, a finding consistent with previous research [33, 34] (see S5). Importantly, our use of DWT to decompose rs-fMRI signals enabled the identification of frequency-specific connectivity changes that correlate with distinct cognitive deficits. For a detailed analysis of how connectivity alterations manifest across different temporal resolutions, from high-frequency to low-frequency bands, see the Supplementary Materials (S6). Recent studies have demonstrated that DWT-derived connectivity metrics not only enhance sensitivity to pathological changes but also correlate strongly with clinical symptom severity and domain-specific impairments in AD and MCI [35, 36]. These results underscore the potential of wavelet-based approaches to provide biomarkers that are both neurophysiologically meaningful and clinically relevant. Overall, the altered functional connectivity patterns highlight disruptions in memory, executive function, and visual processing networks, supporting the robustness and relevance of the findings.

In the second phase of our study, we used graph theoretical measures to accurately distinguish patients with AD, MCI, and healthy controls. The classification accuracy using graph measures of brain networks obtained from original rs-fMRI signals, was 80% for SVM and 95.7895% for the Bayes classifier (Figure S6). There were improvements in the accuracies when using DWT decomposed rs-fMRI signals. Overall, the naïve Bayes classifier provided much better performance than the SVM (Figure 4). From Figure 4 and Tables S4-S7, it can be seen that in most cases, the level 5 decomposition (~0.01–0.02 Hz, given a TR of 3000 ms) resulted in the best classification accuracy among other levels of decomposition for each type of mother wavelets. This suggests that low-frequency components, which reflect stable, long-range brain interactions, are particularly informative for distinguishing these groups. The discriminative power of these low-frequency components underscores the relevance of slow BOLD oscillations as biomarkers of disease-related network alterations.

While all wavelet families performed well, some differences were observed; a detailed comparison of classification performance across different wavelet families is provided in the Supplementary Materials (see S7). Analysis of the top ten most discriminative features revealed a dominance of centrality-based metrics, with repeated involvement of ROI 110 (Lingual_L, DMN), suggesting that the integration and importance of this region within the network is significantly altered across disease stages. The most frequently selected features originated from regions linked to the DMN, Memory Retrieval, Salience, and Subcortical Networks, aligning well with known pathological progression in Alzheimer's disease. Furthermore, an analysis of global network measures across DWT levels showed that networks at lower frequencies were more segregated and locally clustered, while higher-frequency networks emphasized more integrated global communication (Figure S7). A detailed report on the trends of specific global metrics across decomposition levels is available in the Supplementary Materials (see S8).

One limitation of this study lies in the multi-center nature of the ADNI dataset. Although standardized protocols were used, subtle site-specific differences may remain. Future work should consider incorporating statistical harmonization techniques to control for this potential variability. Another limitation is the significant difference in age between the participant groups. Since age can influence brain connectivity, future studies should include age as a covariate to mitigate this potential confounding effect. Finally, Pearson's correlation was used to estimate brain connectivity. While common, it only captures linear relationships. Future studies could explore non-linear measures to better characterize brain networks.

## 5- Conclusions

We used graph theoretical analysis and pattern recognition on resting-state fMRI data to classify normal controls, MCI patients, and AD patients. The method identified key brain regions and connections, particularly within the default mode network and visual areas, that distinguish the groups. Findings were consistent across frequency bands, although some variations highlighted the dynamic nature of brain function. Using DWT-decomposed graphs improved classification accuracy, especially at low frequencies, reflecting stable, long-range brain interactions. Overall, this multi-scale approach enhances both diagnostic performance and understanding of frequency-specific brain alterations in neurodegenerative diseases.

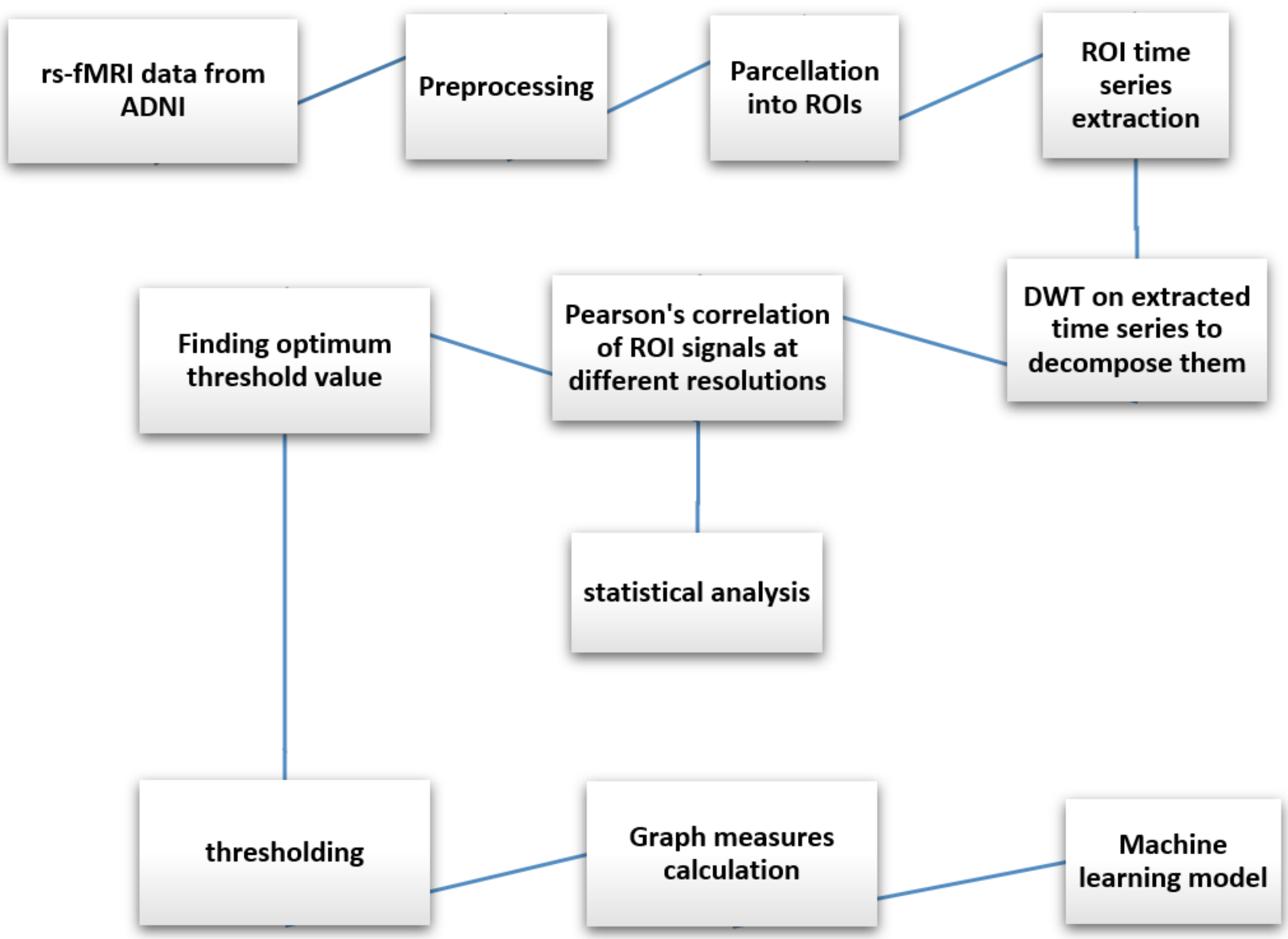

Figure 1: The detailed workflow for the proposed method.

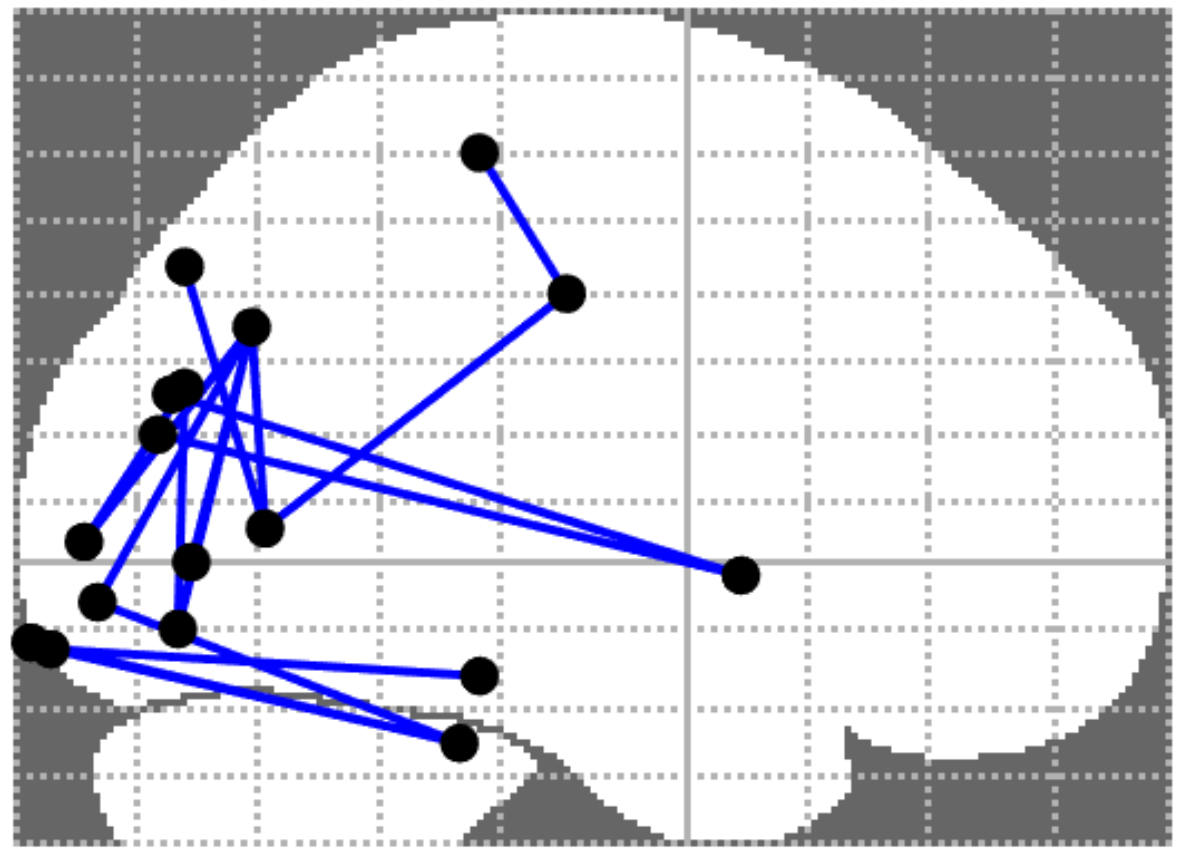

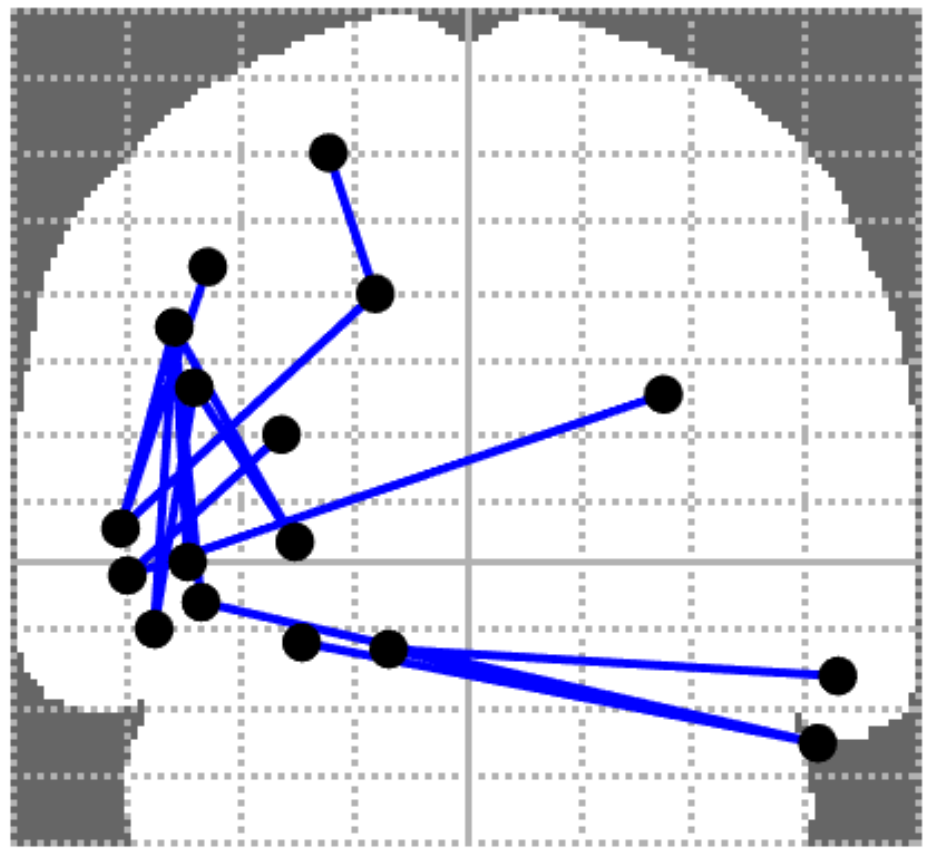

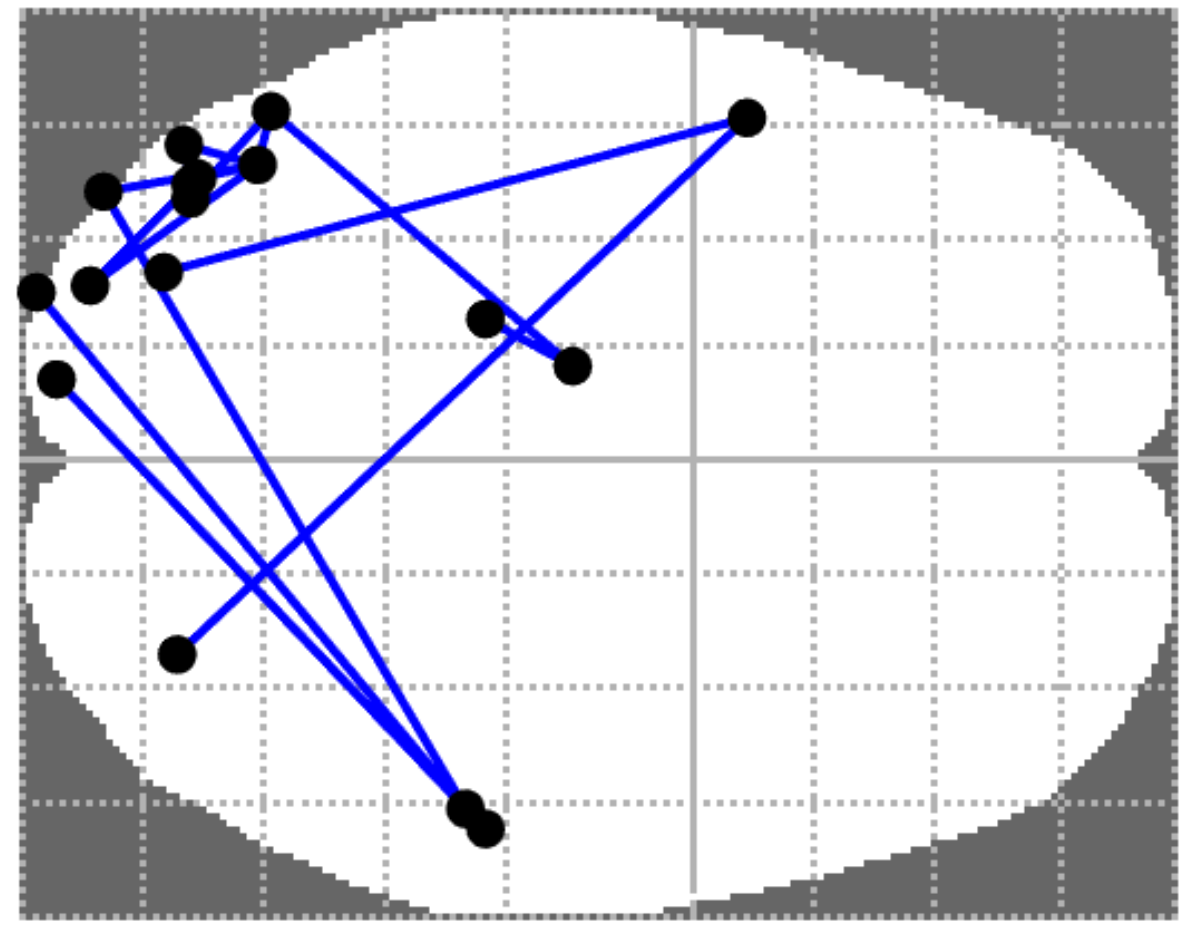

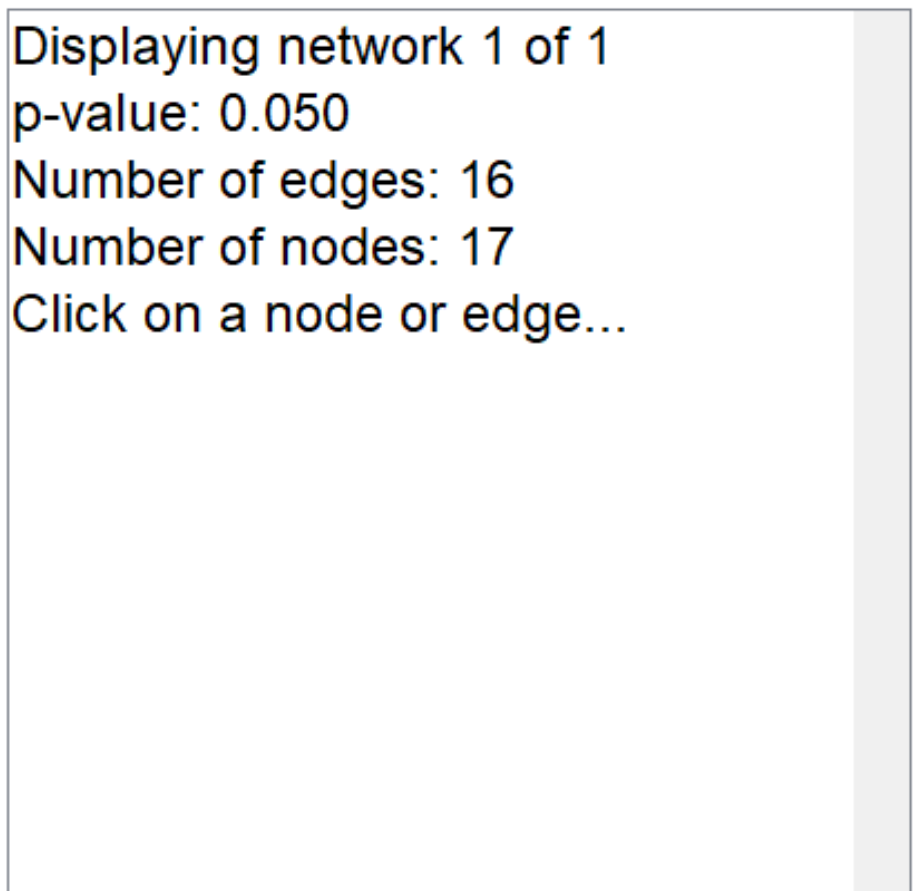

Figure 2: Subnetwork showing significant group differences among HC, MCI, and AD groups identified using the NBS method with an F-test on original rs-fMRI connectivity matrices. Statistical significance was determined via nonparametric permutation testing with FDR correction.

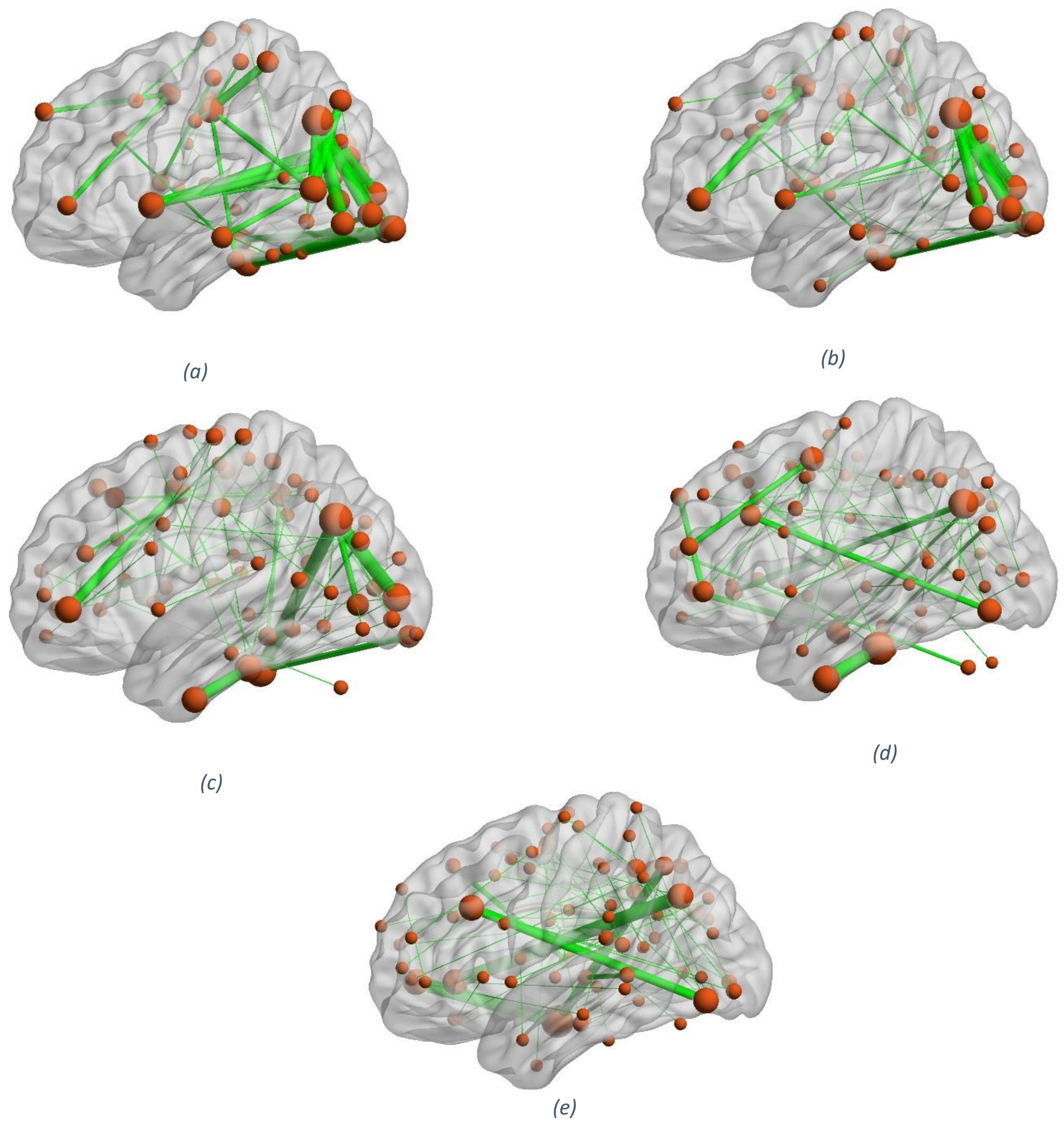

Figure 3: Nodes and edges with significant differences between three groups of HC, MCI, and AD. Approximation coefficients of (a: level1, b: level2, c: level3, d: level4, e: level5) decomposition of rs-fMRI signals were used to construct raw connectivity matrices. The results of NBS analysis for 20 different wavelet types were accumulated. The greater the size of nodes and edges, the more frequent the node and edge. The size of nodes represents the strength of nodes in the accumulated matrix. The size of each edge represents the weight of the corresponding connection in the accumulated matrix (the number of appearances of that edge in NBS analysis on raw connectivity matrices of (a: level1, b: level2, c: level3, d: level4, e: level5) decomposition for 20 different wavelet types). Plots of this figure were created using the BrainNet Viewer software package (http://nitrc.org/projects/bnv/).

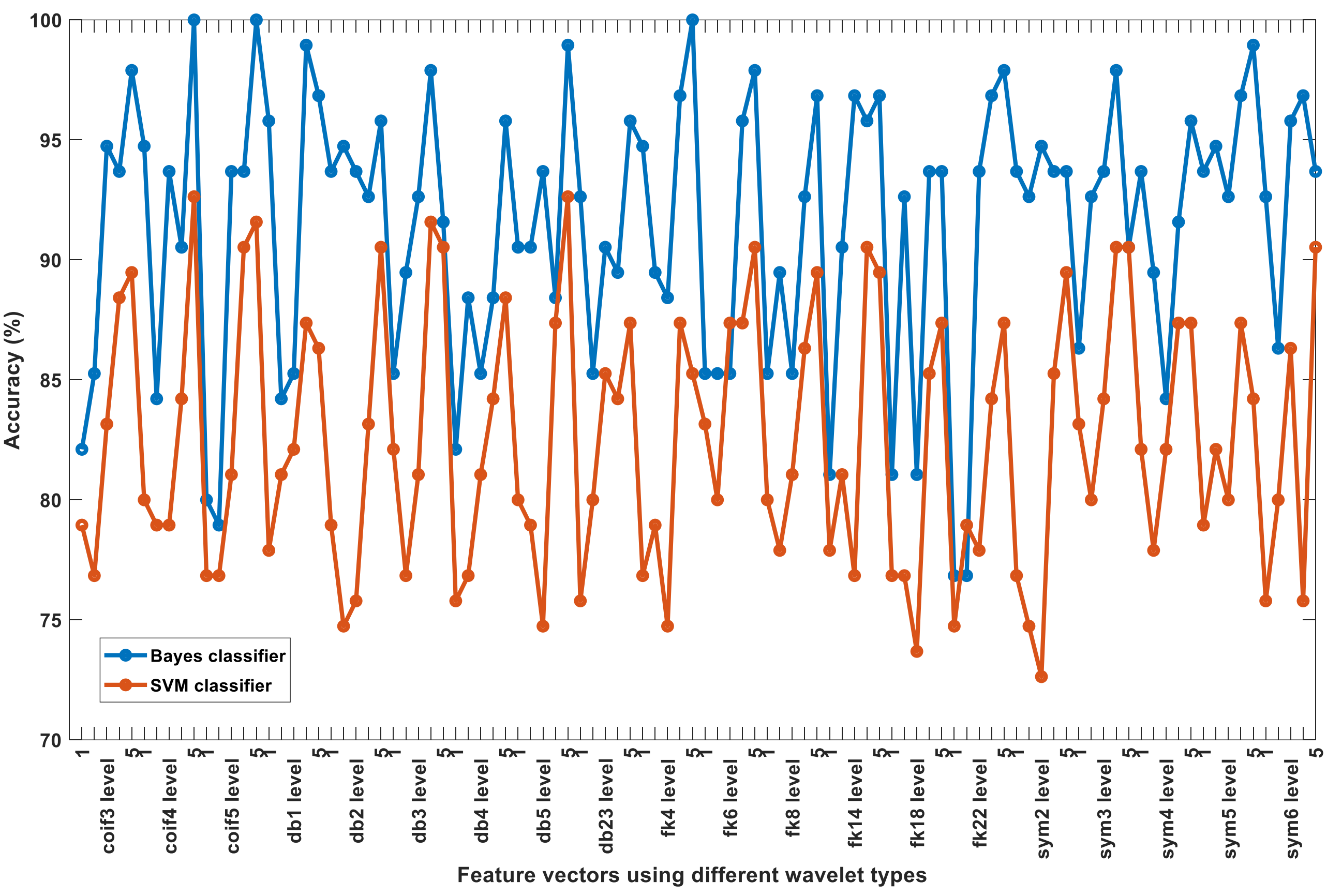

Figure 4: Comparison of performance of naïve Bayes classifier and SVM in classification of three groups of HC, MCI, and AD. Horizonal axis shows 100 different methods of feature extraction used in this study. Twenty different wavelet types at 5 different levels of decomposition were used to decompose original rs-fMRI signals. Decomposed signals were correlated and brain networks were calculated. The graph measures of calculated brain networks were employed as differentiating features.

# Supplementary Materials

## S1: Threshold Selection via Maximum Global Cost Efficiency

The analysis of brain functional networks often necessitates an optimal balance between efficient information processing and the biological cost of connections. This trade-off is quantitatively captured by the concept of global cost efficiency (GCE), which represents the difference between a network's global efficiency—a measure of its capacity for parallel information transfer—and its connection cost, or wiring density. Specifically, GCE is calculated as:

$$GCE = Eglob - C \qquad (1)$$

where Eglob is the global efficiency of the network and C is its cost. he cost C of the network, representing its wiring density, is formalized as:

$$C = \frac{1}{N(N-1)} \sum_{i \neq j \in g} G_{i,j} \qquad (2)$$

Here, N is the total number of nodes in the network, and $G_{i,j}$ is 1 if a connection exists between node i and node j, and 0 otherwise. The term N(N−1) represents the total number of possible directed connections (excluding self-loops).

The theoretical underpinning is that brain networks, like other complex systems, evolve towards an "economical small-world" configuration, characterized by high efficiency at a relatively low connection cost. The provided code systematically explores this trade-off by iteratively constructing binary networks across a range of proportional thresholds. For each threshold, the global efficiency and connection cost are calculated, and their difference yields the GCE. The optimum threshold is then identified as the one that maximizes this GCE, thereby defining a network configuration that is both highly efficient for information flow and parsimonious in its resource allocation, consistent with principles of neurobiological organization [1].

## S2: The Network-Based Statistic (NBS) method

The general procedure involved:

- Calculating a statistical test (e.g., F-statistic) for each individual connection in the connectivity matrix to assess differences between groups.
- Applying an initial primary threshold to these individual connection-wise statistics, retaining only those connections exceeding this threshold.
- Identifying all connected components within the graph formed by these suprathreshold connections.
- Assessing the statistical significance of each identified component's size (number of links) using a non-parametric permutation testing procedure. This involves generating a null distribution of maximal component sizes from a large number of random permutations of group assignments. A component's p-value is then derived by comparing its observed size to this null distribution, thereby controlling the family-wise error rate (FWE) at the component level. The NBS is particularly advantageous for detecting distributed network alterations, as it leverages the topological organization of effects rather than relying solely on the statistical significance of individual links.

In addition to the NBS, the FDR method was utilized to control for multiple comparisons at the individual link level. While the NBS controls the FWE for components, FDR controls the expected proportion of false positives among the declared significant individual connections. This provides a complementary perspective, allowing for assessment of significance both for interconnected subnetworks (via NBS) and for individual connections (via FDR). Both methods were implemented using the NBS toolbox [2].

**S3: Network Alterations Across Wavelet Decomposition Levels**

An investigation of the resting-state networks corresponding to the altered nodes across different levels of wavelet decomposition (as shown in Table S2) revealed the following patterns of disruption:

- First Level: Altered regions were mainly in DMN, visual, uncertain, sensory/somatomotor hand, and dorsal attention networks.
- Second Level: Changes appeared in DMN, visual, uncertain, salience, and sensory/somatomotor hand.
- Third Level: Disruptions were mostly in uncertain, DMN, fronto-parietal task control, visual, and salience networks.
- Fourth Level: Affected regions included DMN, fronto-parietal task control, uncertain, visual, and salience.
- Fifth Level: Most alterations were observed in DMN, fronto-parietal task control, visual, salience, and sensory/somatomotor hand.

**S4: Detailed Analysis of Frequently Disrupted Brain Regions**

An analysis of the second column of Table S2 reveals that the most frequently disrupted regions (across all wavelet types and decomposition levels) included:

- Angular Gyrus (Left) (ROI 99): A key DMN region often disrupted in AD, associated with memory and language deficits [3-5]. Suggested as a target for non-invasive stimulation therapies (e.g., rTMS) [5].
- Temporal Regions (ROIs 26, 250, 147, 64): Inferior and middle temporal gyri and superior temporal pole, involved in memory, semantics, and social/emotional processing [6-8].

- Occipital Regions (ROIs 8, 154, 156, 159, 12): Visual processing areas, indicating visual and visuospatial dysfunction in AD [9, 10].
- Precentral Gyrus (ROI 225): Related to motor control, possibly reflecting broader network disruption [11].
- Middle Frontal Orbital Gyrus (ROI 192): Part of executive networks, associated with decision-making impairments [12].
- Fusiform Gyrus (ROI 5): Important for facial recognition, consistent with visual recognition deficits in AD [13].

## S5: Most Frequently Disrupted Functional Connection

Looking at the third column of Table S3, the most frequently disrupted connection (edge) was between the left angular gyrus and left middle occipital gyrus, linking disruptions in DMN and visual processing pathways. Reduced connectivity between these areas is thought to contribute to memory loss, visuospatial deficits, and overall cognitive decline in AD and MCI, consistent with previous research [14, 15].

## S6: Detailed Analysis of Brain Functional Connectivity Across Temporal Resolutions

An analysis of the rows in Table S2 reveals alterations in brain functional connectivity among AD, MCI, and HC groups across multiple temporal resolutions. At the first level of decomposition, corresponding to high-frequency brain dynamics, the most frequently disrupted regions included the Left Angular Gyrus, Right Inferior Temporal Gyrus, Left Middle Temporal Gyrus, Left Superior Temporal Pole, Left Inferior Occipital Gyrus, Left Middle Occipital Gyrus, and Left Lingual Gyrus, implicating disturbances in rapid sensory, emotional, and visual information processing. At the fifth level of decomposition, capturing low-frequency, sustained brain interactions, disruptions were most prominent in the Right Middle Temporal Gyrus, Right Inferior Frontal Gyrus (Orbital Part), Left Inferior Frontal Gyrus (Triangular Part), and again the Left Inferior Occipital Gyrus, indicating impairments in higher-order cognitive functions such as memory consolidation, decision-making, and executive control. Notably, occipital and temporal regions exhibited significant disruptions across both transient and sustained frequency bands, suggesting that AD and MCI pathology affects both fast, localized processes and slower, integrative cognitive networks.

## S7: Detailed Classification Performance Across Wavelet Families and Decomposition Levels

The classification results across different wavelet families (Coiflet, Daubechies, Fejér-Korovkin, and Symlet), as summarized in Tables S4–S7, demonstrate notable variability in performance. Generally, all wavelets achieved high accuracy with the Naïve Bayes classifier and SVM, but some differences were observed. For instance, Coiflet wavelets (coif3–coif5) at higher decomposition levels yielded accuracy up to 100%, particularly with Naïve Bayes. Daubechies wavelets showed consistently high accuracies as well, with db1 and db5 often outperforming others at various levels. Fejér-Korovkin wavelets presented a wider range of accuracies, with some levels reaching perfect classification (e.g., fk4 level 5). Symlet wavelets also achieved robust accuracies, with sym5 showing especially strong performance at levels 4 and 5. The number of features selected after the Fisher score and FSFS stages varied substantially across wavelets and levels, indicating differences in feature redundancy and relevance. These findings likely reflect the distinct time-frequency localization and smoothness properties of each wavelet family. Overall, smoother wavelets with better frequency localization, such as Coiflets and Symlets, tended to yield higher classification accuracies.

## S8: Analysis of Global Network Topology Across DWT Levels

To explore how graph topology varies across temporal scales, we compared global network measures across the five DWT decomposition levels using the db1 wavelet. As shown in Figure S7, characteristic path length and clustering coefficient both increased from Level 1 to Level 5, indicating a shift toward greater local connectivity and reduced network integration at lower frequencies. Conversely, global efficiency decreased with increasing level, reflecting diminished capacity for long-range information transfer. Assortativity also increased across levels, suggesting enhanced network resilience and preference for connections among similar-degree nodes. Small-worldness remained relatively stable across all decomposition levels.

Table S1: Demographic of healthy subjects, patients with MCI, and patients with AD. MMSE: Mini-Mental State Examination; CDR: Clinical Dementia Rating.

| | Healthy Control | Patients with MCI | Patients with AD | *P*-value |
|---|---|---|---|---|
| Number | 45 | 89 | 34 | - |
| Male/Female | 19/26 | 43/46 | 16/18 | $\chi^2(2) = 0.45$, p = 0.80† |
| Age | 75.9±6.79 | 71.77±7.78 | 72.54±7.02 | $p = 0.006$ (HC vs. MCI), n.s. for others* |
| MMSE score | 28.95±1.56 | 27.56±2.2 | 21. 24±3.37 | $p < 0.001$ (HC vs. AD, MCI vs. AD), $p = 0.009$ (HC vs. MCI)* |
| CDR score | 0.07±0.21 | 0.49±0.17 | 0.92±0.31 | $p < 0.001$ (HC vs. MCI, HC vs. AD, MCI vs. AD)* |

*one-way ANOVA for continuous variables; †Chi-square test

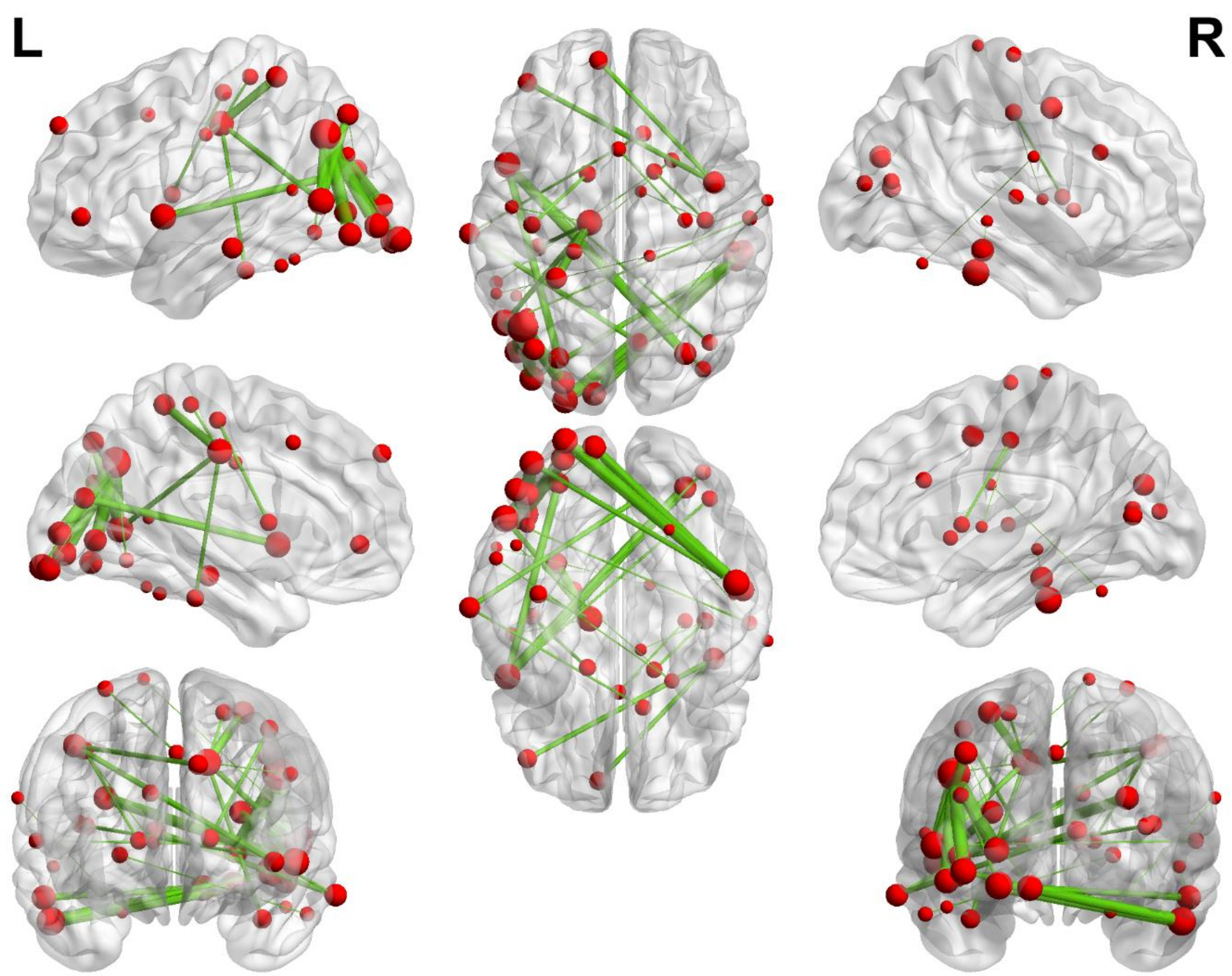

Figure S1: Nodes and edges with significant differences between three groups of HC, MCI, and AD. Approximation coefficients of level 1 decomposition of rs-fMRI signals were used to construct raw connectivity matrices. The results of NBS analysis for 20 different wavelet types were accumulated. The greater the size of nodes and edges, the more frequent node and edge. The size of nodes represents the strength of nodes in the accumulated matrix. The size of each edge represents the weight of the corresponding connection in the accumulated matrix (the number of appearances of that edge in NBS analysis on raw connectivity matrices of level1 decomposition for 20 different wavelet types). Plots of this figure were created using the BrainNet Viewer software package (http://nitrc.org/projects/bnv/).

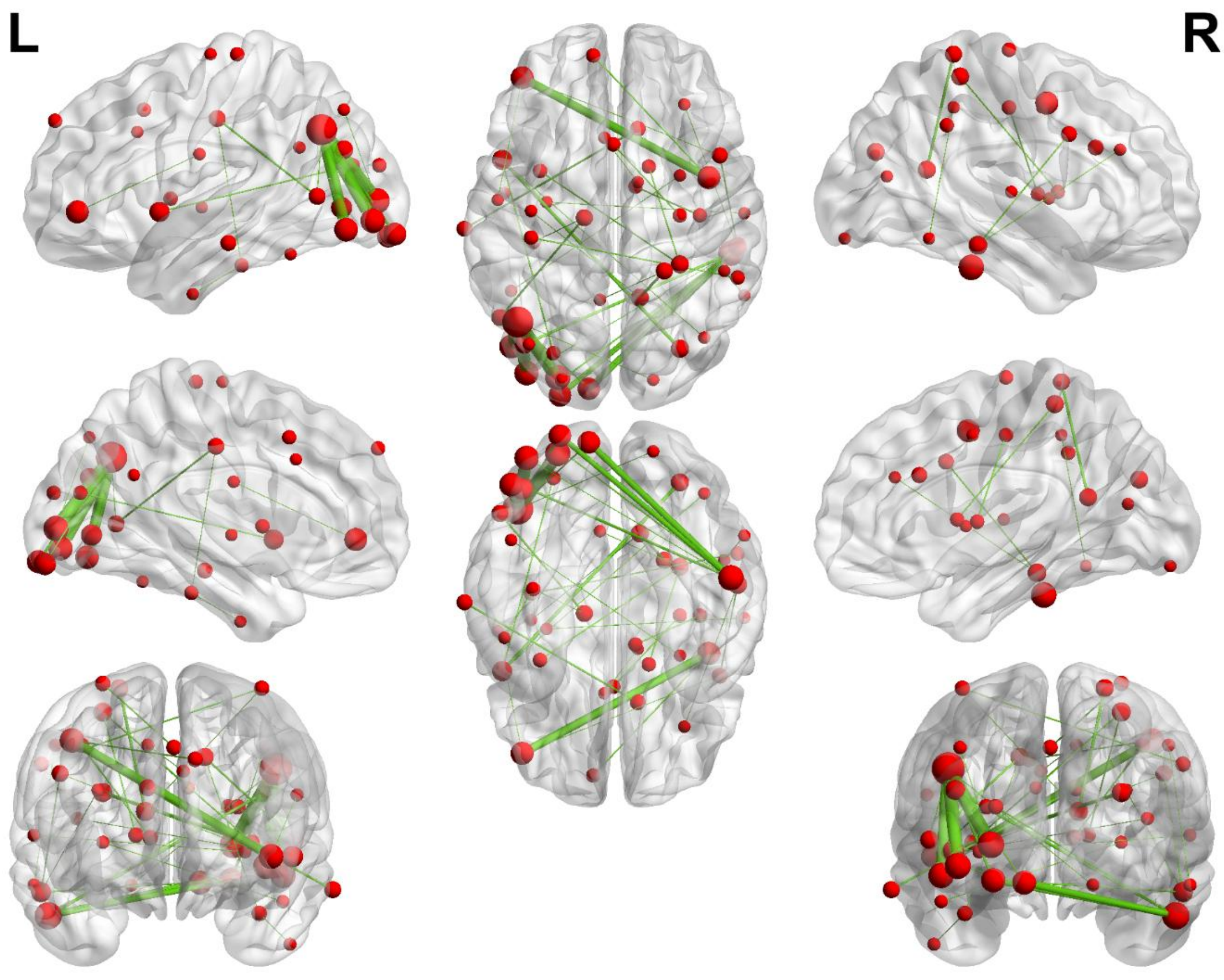

Figure S2: Nodes and edges with significant differences between three groups of HC, MCI, and AD. Approximation coefficients of level 2 decomposition of rs-fMRI signals were used to construct raw connectivity matrices. The results of NBS analysis for 20 different wavelet types were accumulated. The greater the size of nodes and edges, the more frequent node and edge. The size of nodes represents the strength of nodes in the accumulated matrix. The size of each edge represents the weight of the corresponding connection in the accumulated matrix (the number of appearances of that edge in NBS analysis on raw connectivity matrices of level 2 decomposition for 20 different wavelet types). Plots of this figure were created using the BrainNet Viewer software package (http://nitrc.org/projects/bnv/).

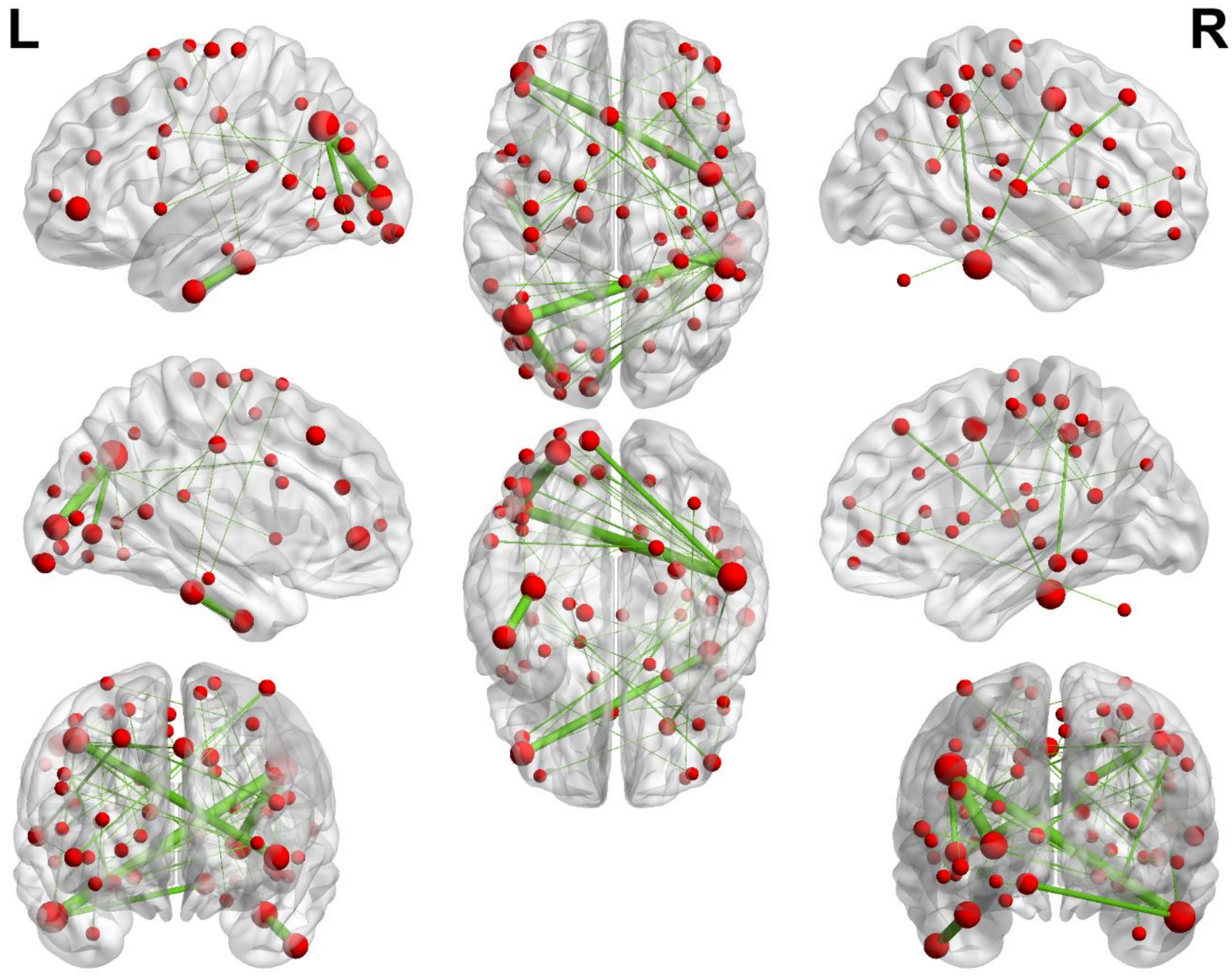

Figure S3: Nodes and edges with significant differences between three groups of HC, MCI, and AD. Approximation coefficients of level 3 decomposition of rs-fMRI signals were used to construct raw connectivity matrices. The results of NBS analysis for 20 different wavelet types were accumulated. The greater the size of nodes and edges, the more frequent node and edge. The size of nodes represents the strength of nodes in the accumulated matrix. The size of each edge represents the weight of the corresponding connection in the accumulated matrix (the number of appearances of that edge in NBS analysis on raw connectivity matrices of level 3 decomposition for 20 different wavelet types). Plots of this figure were created using the BrainNet Viewer software package (http://nitrc.org/projects/bnv/).

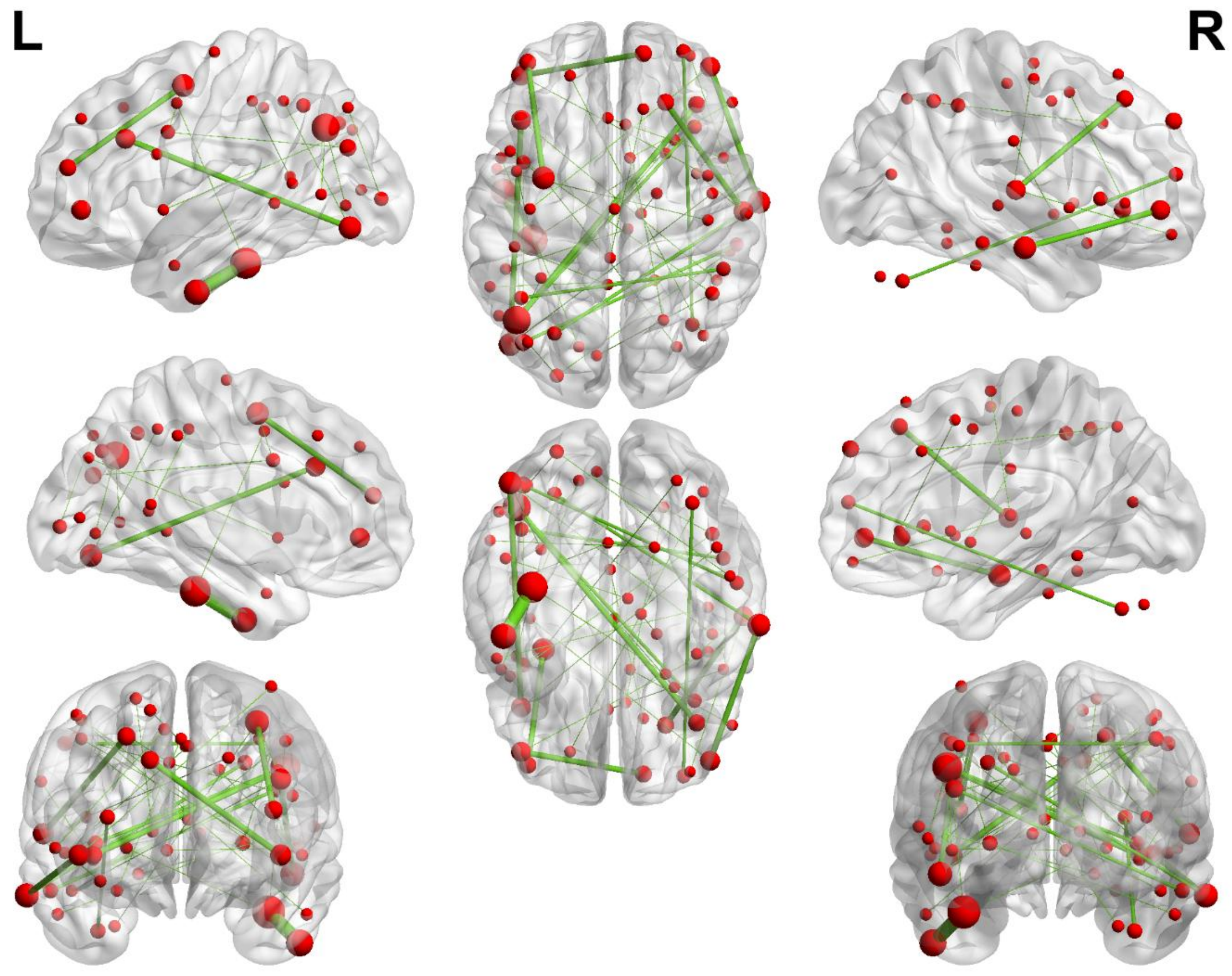

Figure S4: Nodes and edges with significant differences between three groups of HC, MCI, and AD. Approximation coefficients of level 4 decomposition of rs-fMRI signals were used to construct raw connectivity matrices. The results of NBS analysis for 20 different wavelet types were accumulated. The greater the size of nodes and edges, the more frequent node and edge. The size of nodes represents the strength of nodes in the accumulated matrix. The size of each edge represents the weight of the corresponding connection in the accumulated matrix (the number of appearances of that edge in NBS analysis on raw connectivity matrices of level 4 decomposition for 20 different wavelet types). Plots of this figure were created using the BrainNet Viewer software package (http://nitrc.org/projects/bnv/).

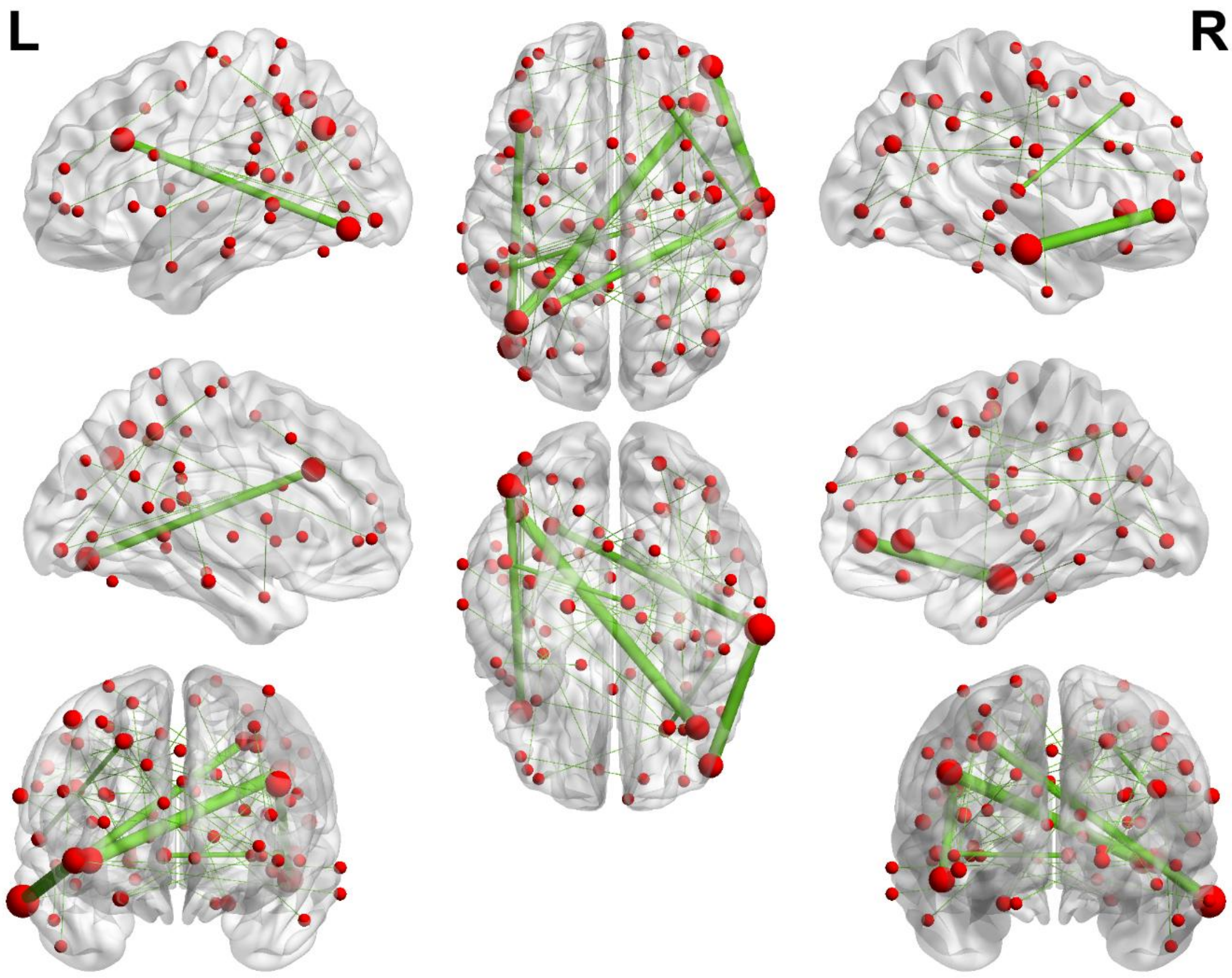

Figure S5: Nodes and edges with significant differences between three groups of HC, MCI, and AD. Approximation coefficients of level 5 decomposition of rs-fMRI signals were used to construct raw connectivity matrices. The results of NBS analysis for 20 different wavelet types were accumulated. The greater the size of nodes and edges, the more frequent node and edge. The size of nodes represents the strength of nodes in the accumulated matrix. The size of each edge represents the weight of the corresponding connection in the accumulated matrix (the number of appearances of that edge in NBS analysis on raw connectivity matrices of level 5 decomposition for 20 different wavelet types). Plots of this figure were created using the BrainNet Viewer software package (http://nitrc.org/projects/bnv/).

Table S2: Significant nodes and connections (edges) in the aggregated NBS matrix. The aggregated NBS matrix was resulted by aggregating 20 sparse matrices of significant connections. Each matrix of significant connection was obtained by applying NBS analysis on connectivity matrices of three groups of HC, MCI, and AD. These connectivity matrices were resulted by correlating decomposed rs-fMRI signals using 20 different wavelet types at specified level of decomposition.

| level of decomposition of rs-fMRI signals. The approximate coefficients of decomposed signals were then correlated to construct connectivity matrices for three groups of HC, MCI, and AD | Most frequent significant nodes across 20 different wavelet types. | | Most frequent significant connections (edges) across 20 different wavelet types. | |
|---|---|---|---|---|
| | ROI No. (from 264 putative functional area atlas) | Strength of nodes in the aggregated NBS matrix | The connection (edge): (ROI1, ROI2) | Weight of connections (edges) in the aggregated NBS matrix |
| level 1 of decomposition | 99 | 103 | (26,8) | 20 |
| | 26 | 43 | (99,154) | 20 |
| | 250 | 41 | (99,156) | 20 |
| | 64 | 36 | (99,159) | 20 |
| | 8 | 34 | (99,155) | 18 |
| | 38 | 33 | (99,250) | 17 |
| | 156 | 28 | (64,179) | 16 |
| | 159 | 23 | (12,26) | 15 |
| | 12 | 23 | (34,38) | 13 |
| | 154 | 20 | (102,250) | 13 |
| | 155 | 18 | (64,158) | 12 |
| | 27 | 18 | (8,27) | 10 |
| | 179 | 16 | | |
| | 225 | 15 | | |
| | 102 | 14 | | |
| | 158 | 13 | | |
| | 34 | 13 | | |
| | 91 | 12 | | |
| level 2 of decomposition | 99 | 74 | (99,159) | 20 |
| | 26 | 25 | (99,156) | 19 |
| | 159 | 23 | (99,154) | 17 |
| | 156 | 20 | (192,225) | 13 |
| | 154 | 17 | (12,26) | 10 |
| | 225 | 15 | (99,155) | 10 |
| | 192 | 15 | (8,26) | 8 |
| | 8 | 14 | (64,179) | 5 |
| | 12 | 13 | | |
| | 155 | 10 | | |
| level 3 of decomposition | 99 | 29 | (99,159) | 11 |
| | 26 | 23 | (99,26) | 10 |
| | 159 | 13 | (3,5) | 9 |
| | 225 | 11 | (192,225) | 8 |
| | 5 | 11 | (12,26) | 5 |
| | 192 | 10 | (99,155) | 5 |
| | 3 | 9 | (88,137) | 3 |
| | | | (138,208) | 3 |
| level 4 of decomposition | 5 | 8 | (3,5) | 6 |
| | 99 | 7 | (88,137) | 3 |
| | 3 | 6 | (99,224) | 3 |
| | 253 | 5 | (132,192) | 3 |
| | 154 | 5 | (147,205) | 3 |
| | 147 | 5 | (154,190) | 3 |
| | | | (210,253) | 3 |
| level 5 of decomposition | 147 | 7 | (99,224) | 4 |
| | 224 | 5 | (147,205) | 4 |
| | 205 | 5 | (147,194) | 3 |
| | 190 | 5 | (154,190) | 3 |
| | 154 | 5 | (88,137) | 2 |
| | 99 | 5 | (96,234) | 2 |

Table S3: Significant nodes and connections (edges) in the sparse matrix of significant connections, obtained by applying NBS analysis on connectivity matrices of three groups of HC, MCI, and AD. The connectivity matrices were resulted by correlating original rs-fMRI signals.

| Significant nodes | | Significant connections (edges) |
|---|---|---|
| ROI No. (from 264 putative functional area atlas) | Strength of nodes in the sparse matrix of significant connections | The connection (edge): (ROI 1, ROI 2) |
| 99 | 5 | (8,26) |
| 250 | 3 | (12,26) |
| 26 | 3 | (12,27) |
| 159 | 2 | (26,156) |
| 156 | 2 | (36,38) |
| 154 | 2 | (38,250) |
| 101 | 2 | (64,158) |
| 64 | 2 | (64,179) |
| 38 | 2 | (99,154) |
| 12 | 2 | (99,155) |
| 8 | 1 | (99,156) |
| 27 | 1 | (99,159) |
| 36 | 1 | (99,250) |
| 155 | 1 | (101,154) |
| 158 | 1 | (101,159) |
| 102 | 1 | (102,250) |
| 179 | 1 | |

Table S4: Classification results using Naïve Bayes classifier and SVM. Input features were calculated based on the graphs of brain network constructed using Coiflet wavelets at five different levels of decomposition. Top 600 features of Fisher score feature selection part was selected and fed into the FSFS algorithm.

| Wavelet name | level | naïve Bayes classifier | | SVM | |
|---|---|---|---|---|---|
| | | Accuracy (%) | No. of selected features (out of 2909 features) | Accuracy (%) | No. of selected features (out of 2909 features) |
| coif3 | 1 | 82.105 | 193 | 78.947 | 246 |
| | 2 | 85.263 | 58 | 76.842 | 373 |
| | 3 | 94.737 | 127 | 83.158 | 35 |
| | 4 | 93.684 | 282 | 88.421 | 182 |
| | 5 | 97.895 | 160 | 89.474 | 179 |
| coif4 | 1 | 94.737 | 188 | 80 | 330 |
| | 2 | 84.211 | 130 | 78.947 | 35 |
| | 3 | 93.684 | 238 | 78.947 | 27 |
| | 4 | 90.526 | 143 | 84.211 | 176 |
| | 5 | 100 | 354 | 92.632 | 129 |
| coif5 | 1 | 80 | 134 | 76.842 | 329 |
| | 2 | 78.947 | 95 | 76.842 | 17 |
| | 3 | 93.684 | 131 | 81.053 | 114 |
| | 4 | 93.684 | 163 | 90.526 | 230 |
| | 5 | 100 | 172 | 91.579 | 277 |

Table S5: Classification results using Naïve Bayes classifier and SVM. Input features were calculated based on the graphs of brain network constructed using Daubechies wavelets at five different levels of decomposition. Top 600 features of Fisher score feature selection part was selected and fed into the FSFS algorithm.

| Wavelet name | level | naïve Bayes classifier | | SVM | |
|---|---|---|---|---|---|
| | | Accuracy (%) | No. of selected features (out of 2909 features) | Accuracy (%) | No. of selected features (out of 2909 features) |
| db1 | 1 | 95.789 | 108 | 77.895 | 348 |
| | 2 | 84.211 | 252 | 81.053 | 401 |
| | 3 | 85.263 | 341 | 82.105 | 152 |
| | 4 | 98.947 | 263 | 87.368 | 191 |
| | 5 | 96.842 | 177 | 86.316 | 216 |
| db2 | 1 | 93.684 | 169 | 78.947 | 339 |
| | 2 | 94.737 | 153 | 74.737 | 547 |
| | 3 | 93.684 | 154 | 75.789 | 493 |
| | 4 | 92.632 | 140 | 83.158 | 72 |
| | 5 | 95.789 | 198 | 90.526 | 153 |
| db3 | 1 | 85.263 | 108 | 82.105 | 93 |
| | 2 | 89.474 | 80 | 76.842 | 109 |
| | 3 | 92.632 | 137 | 81.053 | 12 |
| | 4 | 97.895 | 134 | 91.579 | 221 |
| | 5 | 91.579 | 407 | 90.526 | 92 |
| db4 | 1 | 82.105 | 132 | 75.789 | 224 |
| | 2 | 88.421 | 80 | 76.842 | 358 |
| | 3 | 85.263 | 220 | 81.053 | 27 |
| | 4 | 88.421 | 133 | 84.211 | 389 |
| | 5 | 95.789 | 178 | 88.421 | 236 |
| db5 | 1 | 90.526 | 100 | 80 | 101 |
| | 2 | 90.526 | 112 | 78.947 | 194 |
| | 3 | 93.684 | 157 | 74.737 | 509 |
| | 4 | 88.421 | 132 | 87.368 | 189 |
| | 5 | 98.947 | 174 | 92.632 | 150 |
| db23 | 1 | 92.632 | 83 | 75.789 | 450 |
| | 2 | 85.263 | 234 | 80 | 20 |
| | 3 | 90.526 | 167 | 85.263 | 133 |
| | 4 | 89.474 | 366 | 84.211 | 79 |
| | 5 | 95.789 | 317 | 87.368 | 93 |

Table S6: Classification results using Naïve Bayes classifier and SVM. Input features were calculated based on the graphs of brain network constructed using Fejér-Korovkin wavelets at five different levels of decomposition. Top 600 features of Fisher score feature selection part was selected and fed into the FSFS algorithm.

| Wavelet name | level | naïve Bayes classifier | | SVM | |
|---|---|---|---|---|---|
| | | Accuracy (%) | No. of selected features (out of 2909 features) | Accuracy (%) | No. of selected features (out of 2909 features) |
| fk4 | 1 | 94.737 | 133 | 76.842 | 542 |
| | 2 | 89.474 | 141 | 78.947 | 366 |
| | 3 | 88.421 | 162 | 74.737 | 428 |
| | 4 | 96.842 | 273 | 87.368 | 118 |
| | 5 | 100 | 186 | 85.263 | 239 |
| fk6 | 1 | 85.263 | 200 | 83.158 | 84 |
| | 2 | 85.263 | 88 | 80 | 548 |
| | 3 | 85.263 | 105 | 87.368 | 93 |
| | 4 | 95.789 | 123 | 87.368 | 123 |
| | 5 | 97.895 | 185 | 90.526 | 170 |
| fk8 | 1 | 85.263 | 95 | 80 | 186 |
| | 2 | 89.474 | 80 | 77.895 | 346 |
| | 3 | 85.263 | 152 | 81.053 | 31 |
| | 4 | 92.632 | 191 | 86.316 | 54 |
| | 5 | 96.842 | 119 | 89.474 | 243 |
| fk14 | 1 | 81.053 | 75 | 77.895 | 171 |
| | 2 | 90.526 | 198 | 81.053 | 338 |
| | 3 | 96.842 | 180 | 76.842 | 442 |
| | 4 | 95.789 | 197 | 90.526 | 249 |
| | 5 | 96.842 | 109 | 89.474 | 162 |
| fk18 | 1 | 81.053 | 107 | 76.842 | 327 |
| | 2 | 92.632 | 60 | 76.842 | 345 |
| | 3 | 81.053 | 324 | 73.684 | 311 |
| | 4 | 93.684 | 100 | 85.263 | 266 |
| | 5 | 93.684 | 230 | 87.368 | 174 |
| fk22 | 1 | 76.842 | 182 | 74.737 | 5 |
| | 2 | 76.842 | 140 | 78.947 | 346 |
| | 3 | 93.684 | 100 | 77.895 | 31 |
| | 4 | 96.842 | 108 | 84.211 | 207 |
| | 5 | 97.895 | 422 | 87.368 | 252 |

Table S7: Classification results using Naïve Bayes classifier and SVM. Input features were calculated based on the graphs of brain network constructed using Symlet wavelets at five different levels of decomposition. Top 600 features of Fisher score feature selection part was selected and fed into the FSFS algorithm.

| Wavelet name | level | naïve Bayes classifier | | SVM | |
|---|---|---|---|---|---|
| | | Accuracy (%) | No. of selected features (out of 2909 features) | Accuracy (%) | No. of selected features (out of 2909 features) |
| sym2 | 1 | 93.684 | 110 | 76.842 | 345 |
| | 2 | 92.632 | 179 | 74.737 | 479 |
| | 3 | 94.737 | 194 | 72.632 | 494 |
| | 4 | 93.684 | 139 | 85.263 | 137 |
| | 5 | 93.684 | 319 | 89.474 | 225 |
| sym3 | 1 | 86.316 | 170 | 83.158 | 168 |
| | 2 | 92.632 | 103 | 80 | 426 |
| | 3 | 93.684 | 117 | 84.211 | 41 |
| | 4 | 97.895 | 128 | 90.526 | 198 |
| | 5 | 90.526 | 233 | 90.526 | 309 |
| sym4 | 1 | 93.684 | 133 | 82.105 | 13 |
| | 2 | 89.474 | 213 | 77.895 | 433 |
| | 3 | 84.211 | 200 | 82.105 | 25 |
| | 4 | 91.579 | 153 | 87.368 | 124 |
| | 5 | 95.789 | 157 | 87.368 | 72 |
| sym5 | 1 | 93.684 | 127 | 78.947 | 16 |
| | 2 | 94.737 | 161 | 82.105 | 27 |
| | 3 | 92.632 | 130 | 80 | 15 |
| | 4 | 96.842 | 123 | 87.368 | 186 |
| | 5 | 98.947 | 124 | 84.211 | 224 |
| sym6 | 1 | 92.632 | 65 | 75.789 | 558 |
| | 2 | 86.316 | 142 | 80 | 362 |
| | 3 | 95.789 | 136 | 86.316 | 205 |
| | 4 | 96.842 | 143 | 75.789 | 534 |
| | 5 | 93.684 | 159 | 90.526 | 147 |

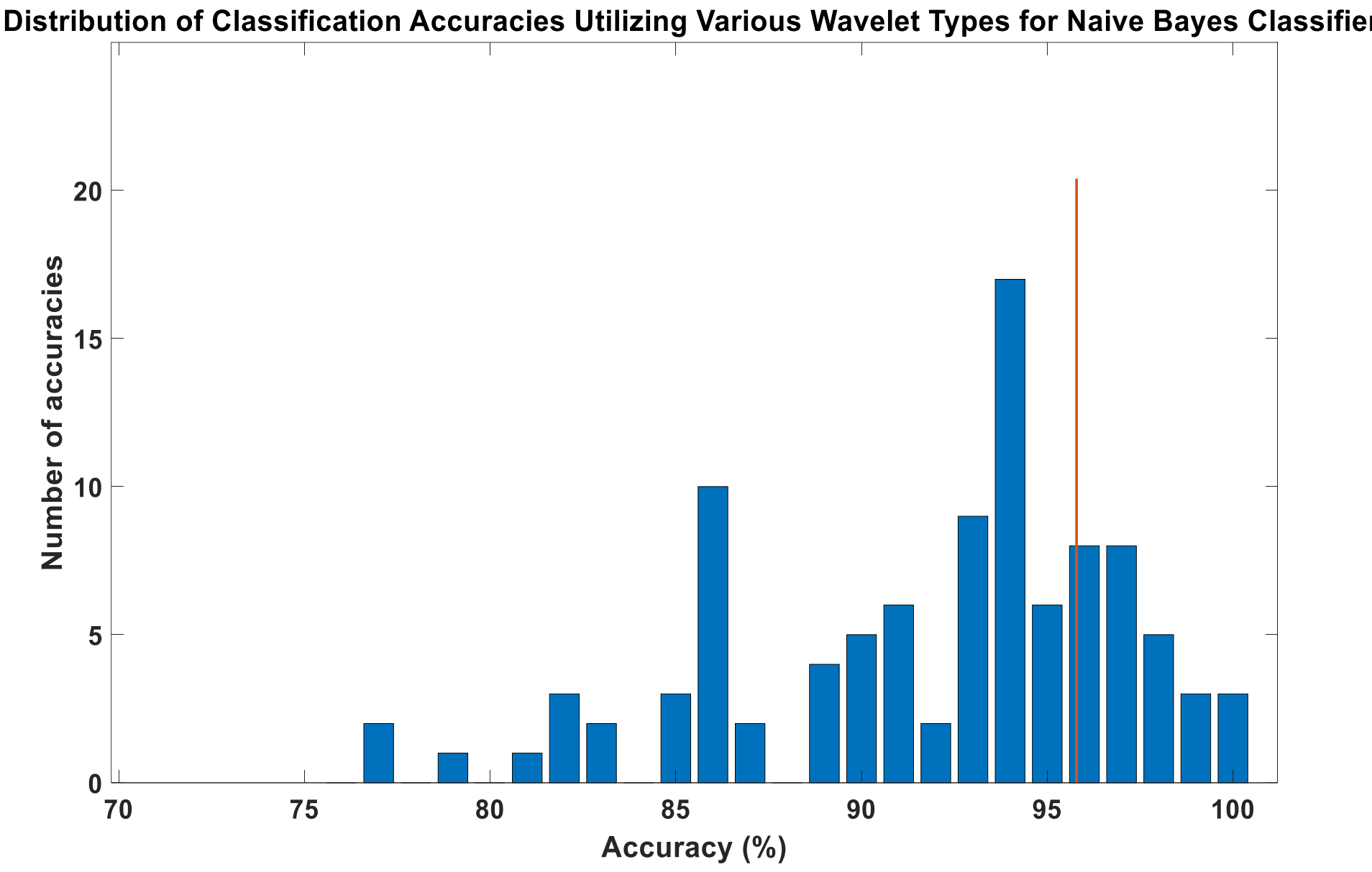

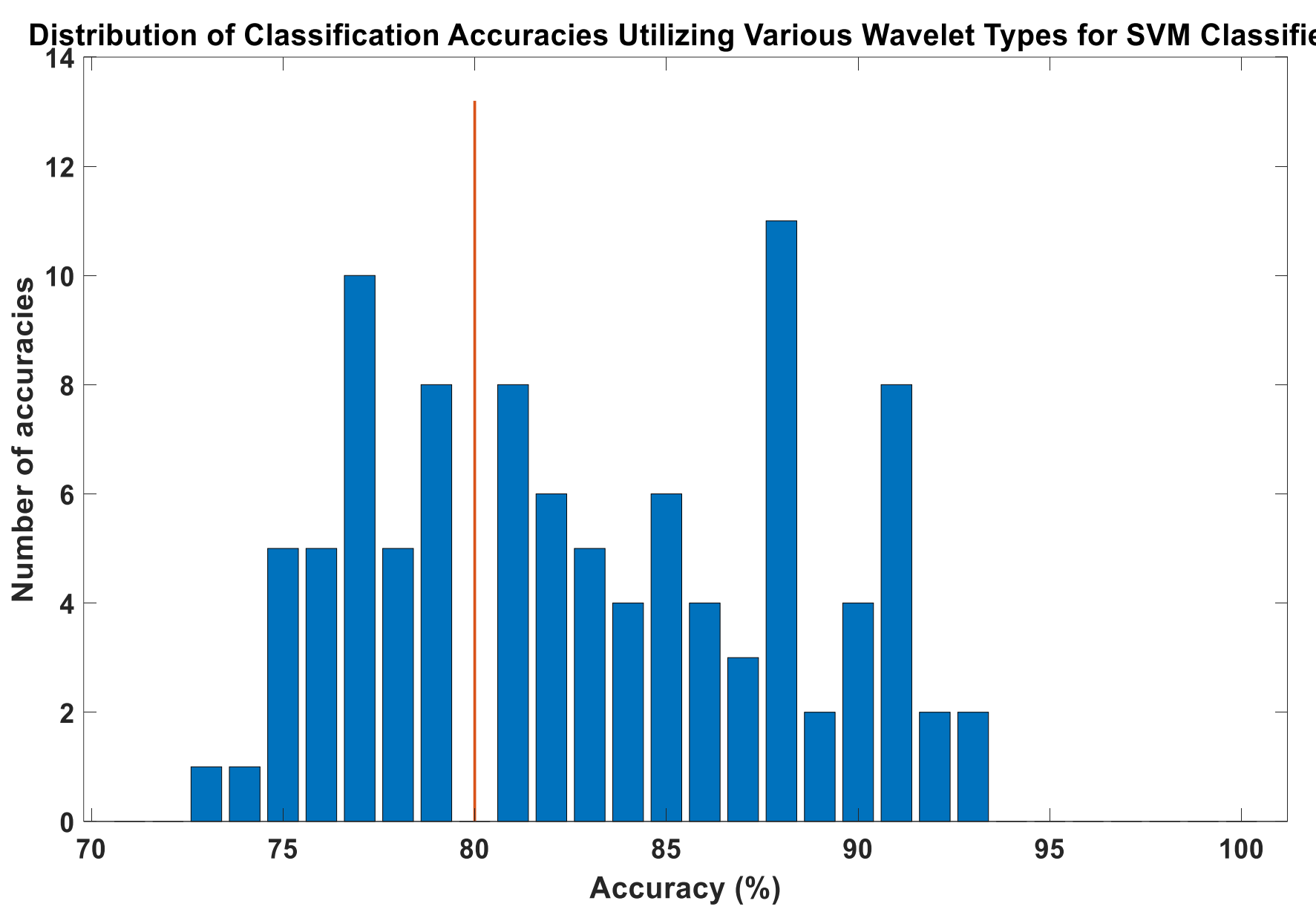

Figure S6: Distribution of classification accuracies over different methods of feature extraction. Top Figure: The data of "Naïve Bayes" column within Tables S4-S7 were used in this graph. The red line shows the Naïve Bayes classification accuracy using graph measures of brain network obtained by correlating original rs-fMRI signals. Bottom Figure: The data of "SVM" column within Tables S4-S7 were used in this graph. The red line shows the SVM classification accuracy using graph measures of brain network obtained by correlating original rs-fMRI signals.

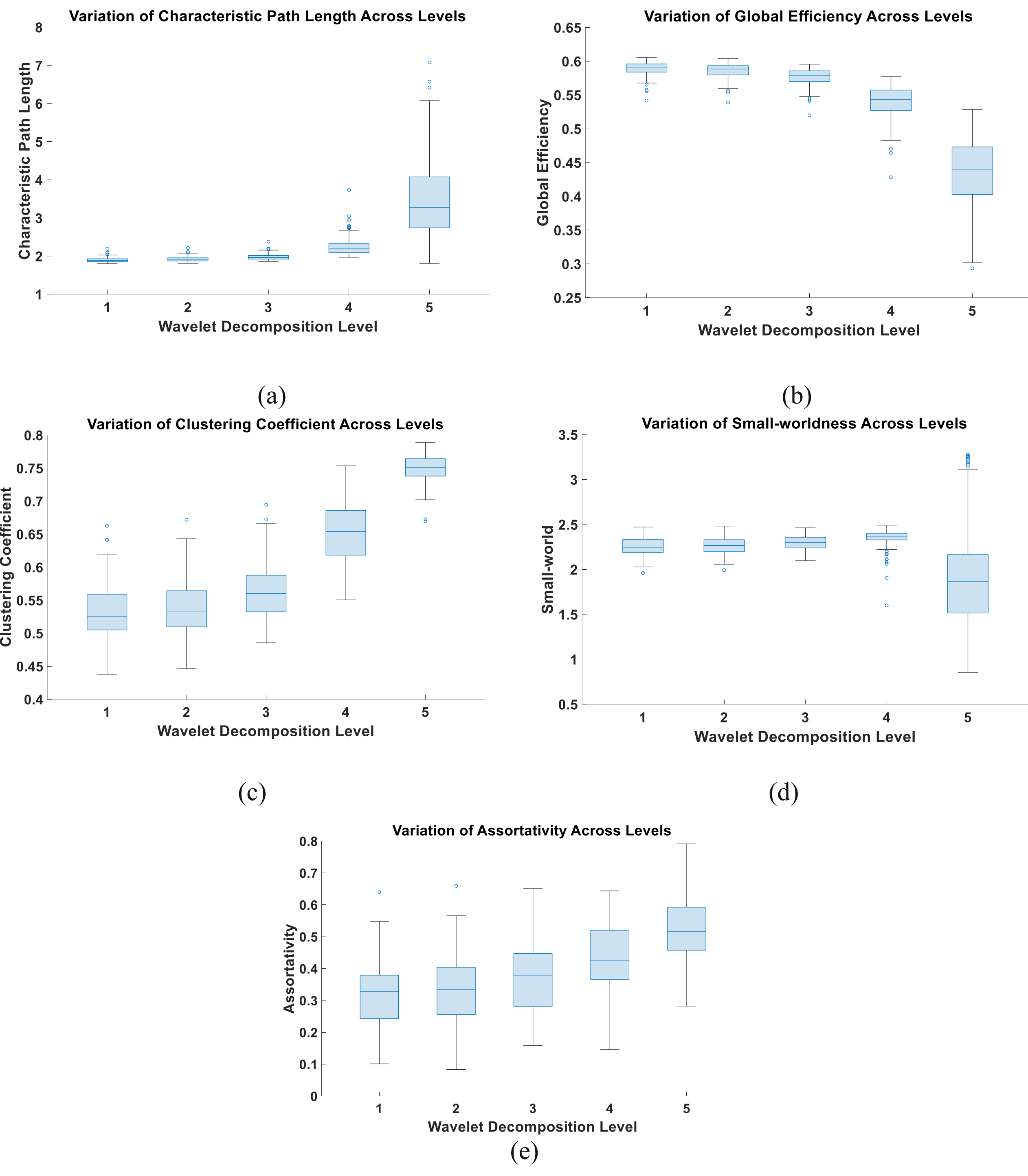

Figure S7: Comparison of global measures across wavelet decomposition levels. Box plots show the distribution of a) characteristic path length, b) global efficiency, c) clustering coefficient, d) small-worldness, and e) assostativity values of all subjects across different levels (Levels 1 to 5) for db1 wavelet type.